\documentclass[article]{revtex4}
\usepackage{pdfpages}
\usepackage{epsf}
\usepackage{subfigure}
\usepackage{amsmath}
\usepackage{eqnarray,amsmath}
\usepackage{epstopdf}
\usepackage{mathrsfs}
\usepackage{natbib}
\usepackage{amssymb,latexsym}
\usepackage{dcolumn}
\usepackage{graphicx}

\makeatletter
\newcommand*{\rom}[1]{\expandafter\@slowromancap\romannumeral #1@}
\makeatother
\def\be{\begin{equation}}
\def\ee{\end{equation}}
\def\ba{\begin{eqnarray}}
\def\ea{\end{eqnarray}}

\begin{document}
\title{Particle creation rate for dynamical black holes}

\author{Javad T. Firouzjaee${}^{1,2}$}
\affiliation{${}^1$School of Astronomy, Institute for Research in Fundamental Sciences (IPM), P. O. Box 19395-5531, Tehran, Iran }
\affiliation{${}^2$Department of Physics (Astrophysics), University of Oxford, Keble Road, Oxford OX1 3RH, UK }
 \email{j.taghizadeh.f@ipm.ir}
\author{George F R Ellis${}^{3}$}
\affiliation{${}^3$Mathematics and Applied Mathematics Department, University of Cape Town, Rondebosch, Cape Town 7701, South Africa}
 \email{gfrellisf@gmail.com}

\begin{abstract}
\textbf{Abstract} \textit{We present the particle creation probability rate around a general black hole as an outcome of quantum fluctuations. Using the uncertainty principle for these fluctuation, we derive a new ultraviolet frequency cutoff for the  radiation spectrum of a dynamical black hole. Using this frequency cutoff, we define the probability creation rate function 
for such black holes. We consider a dynamical Vaidya model, and calculate the probability creation rate for this case 
when its horizon is in a slowly evolving phase. Our results show that one can expect the usual Hawking radiation emission process in the case of a dynamical black hole when it has a slowly evolving horizon. Moreover,  calculating the probability rate for a dynamical black hole gives a measure of when Hawking radiation can be killed off by an incoming flux of matter or radiation. Our result strictly suggests that we have to revise the Hawking radiation expectation for primordial black holes that have grown substantially since they were created in the early universe.  We also infer that this frequency cut off can be a parameter that shows the primordial black hole growth at the emission moment. }
\end{abstract}
%
%

\maketitle

\tableofcontents
\section{Introduction}
Most calculations of the particle creation rate from black holes assume a stationary geometry. However in an astrophysical context, matter and radiation will be falling into the black hole \cite{cosmological black hole, Musco:2004ak}.  This has various consequences. Firstly, the usual black hole uniqueness theorems, which depend  on a vacuum condition, will no longer hold, as recently pointed out by Hawking et al. \cite{Hawking:2016msc}.  Secondly, the usual calculations of black body radiation emission need to be amended to take the dynamic geometry into account (they almost all, explicitly or implicitly, rely on the Killing vectors in the vacuum, which are references for defining the mode functions). For the extended maximal geometry including a bifurcate Killing horizon \cite{Boy69}, one can derive the Hawking radiation formula according to the Hawking-Hartle vacuum, which uses the Killing vector field to define positive and negative mode functions. These definitions implicitly rely on asymptotic flatness of spacetime, which does not hold in the case of a cosmological black hole \cite{cosmological black hole}. \\

As discussed in two previous papers \cite{firouzjaee-ellis14,firouzjaee-ellis15}, Hawking radiation will be suppressed in dynamical black holes with matter or radiation accretion. This will be the case until  very late times, when the dynamical horizon becomes a slowly evolving horizon. It has been claimed that this suppression effect will be negligible for dynamical black holes, but the argument has been made either on the basis of stationary models, or on the basis of scattering models which do not necessarily apply when infalling (positive density) radiation makes the apparent horizon spacelike \cite{ellis13}. We will model dynamic horizons in this paper, but the calculations  are limited to the slowly evolving phase of a black hole. We show that the dynamical effect is negligible in this case (one might assume this is the case on physical grounds, but a calculation is needed to confirm this intuition). But nothing ensures that primordial black holes are slowly evolving, because the background thermal radiation provides the matter flux affecting this evolution and hence affecting Hawking 
radiation. We prove we can neglect Hawking Radiation in this case by calculating the probability creation rate. However we do not estimate timescales. \\

To quantify the particle creation rate for a general black hole, we have to revise the analysis more fundamentally than using 
methods that rely on Killing vectors. It is known that particle creation is a quantum effect which comes from quantum fluctuations. The basis of quantum fluctuations is Heisenberg's uncertainty principle, which somehow violates  energy conservation at the quantum level (and for short times). The virtual created particle-antiparticle pairs can become real in special environments like strong electric fields (Schwinger  effect), special boundaries for the field (Casimir effect), and a black hole event horizon (Hawking radiation). In this  paper we derive the nature of particle creation around a general dynamical black hole according to the quantum fluctuations.
Along the way, we calculate the probability rate of radiation for the general dynamical case, and give an upper bound for the energy spectrum of the radiation. For dynamical black holes, the spectrum range is smaller than in the non-dynamical case. We show that for a slowly evolving black hole the matter flux can be negligible and the black hole radiation luminosity is very close to that in the Schwarzschild (stationary) case. This is not surprising, but one needs a calculation to prove the black hole radiation is the same in these two cases.
\\
 
Section II reviews the motivation for deriving a general picture of Hawking radiation.  The main new results here are, in Section \ref{sec:3}, we describe Hawking radiation according to particle creation in quantum field theory for a general black hole, generalizing the particle picture in a stationary black hole. Then in Section \ref{sec:vaidya}, using the Vaidya exact dynamical model, we calculate the probability rate for particle creation for some dynamical black holes with different matter fluxes. In Sections V and VI we apply this to galactic black holes and primordial black holes respectively. In the latter case there can be a significant reduction of Hawking radiation. \\

We do not specifically investigate different Hawking radiation methods or contexts, but rather consider generic constraints on the radiation spectrum that apply in all cases. We illustrate them by some comments on the case of primordial black holes, when the matter flux is not negligible, and where the results might be very different from what is usually expected. To determine whether this is so or not will required detailed calculation of accretion of matter and radiation in the context of the early universe. The result is not obvious; we have to obtain clarity by investigating dynamical black holes, not the Schwarzschild case. We show how to approach it in the case of genuinely dynamical black holes.

\section{Black hole particle creation}
Most Hawking radiation calculations apply only in the case of stationary geometries (Section \ref{Hstat}), when timelike Killing vectors are present and the apparent and event horizons are identical. However there are some methods that apply in the dynamic case (Section \ref{nonstat}), when either the apparent horizon is timelike and lies outside the event horizon, which is the case of Hawking radiation backreaction in an asymptotically flat context, or is spacelike and lies inside the event horizon, which is the case of cosmological black holes (with matter and infalling radiation) that is of interest to us.

\subsection{Stationary case}\label{Hstat}
The standard way to calculate the particle creation rate is by using the expectation value of the number operator, $N=a^{\dagger}a$, with an appropriate vacuum. The computation involves Bogoliubov transformations relating the past vacuum and future vacuum states.  Quantizing the field, solutions of the classical equations can be written as a linear combination of positive-frequency and negative-frequency parts corresponding  respectively to particles and  antiparticles.\\

To do this calculation we need to assume a stationary geometry so that we can define the positive and negative mode functions by using the timelike Killing vectors. In this way, the Euclidean action method based on a path integral \cite{Gibbons:1977mu} and gravitational anomaly methods \cite{Christensen:1977jc} have been used to calculate Hawking radiation. They both  are  applicable in the case of a  stationary geometry because a key role is played by the exponential relation between an affine parameter and a Killing vector parameter (\cite{Haw75}:(2.16)).  The observed radiation is related to the event horizon, because it must escape to infinity: we have to locate a boundary so that created particles can get to a distant  observer, while trapped particles cannot.  In the stationary case, the apparent horizon and event horizon are the same.

\subsection{The Non-Stationary Case}\label{nonstat}
Consider two rather different non-stationary cases. \\

\textbf{Hawking radiation backreaction in an asymptotically flat context}   
There is  a scattering matrix method for determining the thermal spectrum  that applies in this situation \cite{'tHooft:1996tq}. One can use this method to determine  back-reaction effects on the Schwarzschild metric with fast moving particles. However it seems to 
be hard to build the density matrix of cosmological black holes by this method in practice.\\

\textbf{Hawking radiation for black holes in a cosmological  context}
Recently, the tunneling method was developed to express particle creation events with minimal conditions \cite{pady2000, parikh2000, visser03}. In the particle picture of radiation for a stationary black hole, the tunneling method clarifies that both particle and  antiparticle tunneling contributes to the rate for the Hawking process, where the particle will be outside the  horizon and the antiparticle will be inside the horizon. In terms of a classical interpretation, this means the black hole tidal force near the horizon is so strong that it does not let the created particle pair annihilate each other. But  this picture is not complete because a quantum particle can be in the classically forbidden region. Also most of the Hawking radiation has
	wavelengths of order $m$, and it is only well away from the black hole that one can identify outgoing Hawking ``particles'' unambiguously. The semi-classical energy-momentum tensor looks like a flux of ``real'' 
	positive energy particles only at $r > 3m$ at best, and near the horizon has an ingoing flow of negative energy with no simple particle interpretation. For the most part creation of the Hawking radiation is a highly nonlocal process that cannot be described as taking place in a locally flat region around a single observer. However one can obtain valid local relations by determining the expectation value of the matter stress tensor 
	 at each point.\\

 The best way to describe the quantum particle is by the wave function amplitude, not the tidal force work to create them. This description was used in the tunneling method calculations of \cite{pady2000,parikh2000}. Quantitatively, one can describe  quantum particle creation by using the matter flux $\mathcal{F}$ and energy density $\mathcal{E}$  of radiation determined from the stress tensor expectation value $<T_{\mu\nu}>$ \cite{Chakraborty:2015nwa, firouzjaee-ellis15}: 
\begin{equation}
\mathcal{F}=-u^\mu n^\nu <T_{\mu\nu}>\,,\,\,\,\mathcal{E}=u^\mu u^\nu <T_{\mu\nu}>.
\end{equation} This shows pure created matter properties relative to an  observer with four-velocity $u^\mu$, where $n_\mu$ is a spacelike normal direction to this velocity, $n_\mu u^\mu=0$. These quantities are observer dependent and for both an  infalling observer and static observer are positive \cite{Chakraborty:2015nwa} so that it is meaningless to distinguish a place where the Hawking radiation originates \cite{Gid15}. This method shows that the apparent horizon plays a key role in the particle creation process \cite{hajicek,tunnelingbh,Cli08}. Therefore, this method inherently can be extended to the case of dynamical black holes.  
However, it requires an important assumption: that the  essential adiabatic condition (eikonal approximation) must be satisfied for the wave equation around the apparent horizon, which is related to redshifting and amplification of waves. \\

Reference  \cite{firouzjaee-ellis14} showed  the Shankaranarayanan, Padmanabhan and Visser method (wave interpretation) and the Parikh and Wilczek method (particle method)  are the same in this context, and are in fact both aspects of the tunneling method. It was shown that this  adiabatic condition gives important limitations on which cosmological black holes can emit Hawking radiation, and described the range of the resulting mass spectrum, depending on the flux. Considering the created matter flux and densities, Padmanabhan \cite{Chakraborty:2015nwa} confirms  our analysis that the matter flux and density is positive outside the horizon for both static and infalling observers. Moreover we can see 
 created particles with all different wavelengths outside the horizon according to the tunneling method, but their observed rates are different. \\

\begin{figure}[h]
	\begin{center}
		\includegraphics[scale = 0.34]{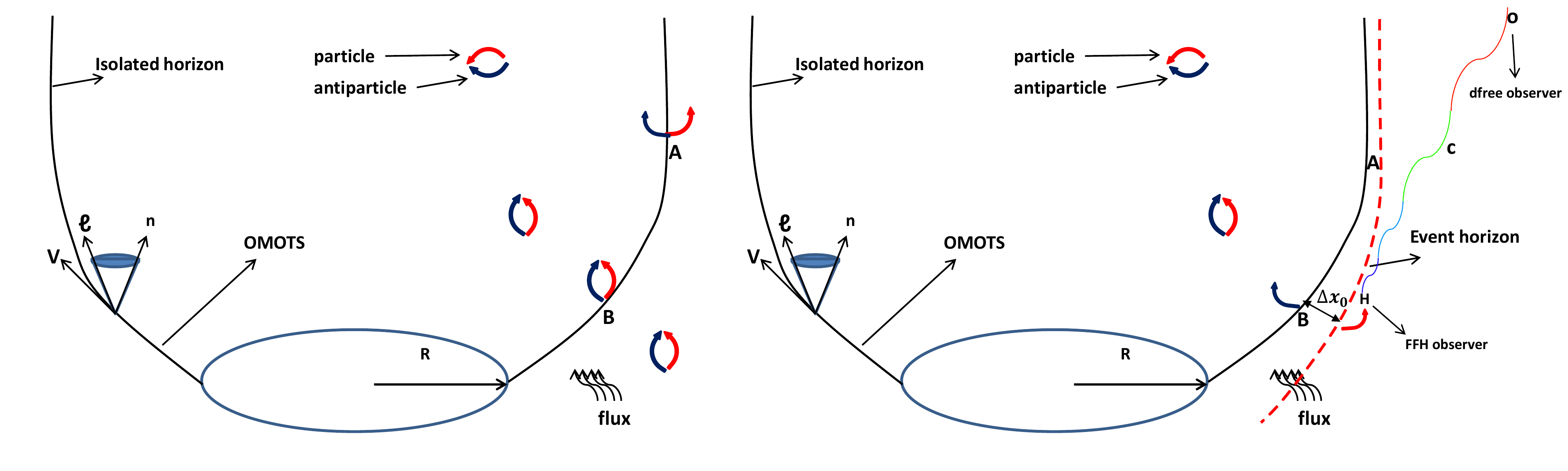}
		\hspace*{10mm} \caption{ \label{particle}
			\textit{On the left hand: Particle creation without the UP. On the right: Particle creation  considering the UP. The OMOTS (Outer Marginally Trapped 3-Surface) is the same apparent horizon in both figures.}}
	\end{center}
\end{figure}
There is another approach which shows  by using the energy conservation principle that a dynamical black hole cannot emit radiation, as described in \cite{firouzjaee-ellis15}. An observer with four-velocity $u^\mu$ can measure the particle energy as
\be \label{energy}
 E=- p_\mu u^\mu
\ee
where $p^\mu$ is the particle four-momentum.
As depicted in the left hand side of  Fig.(\ref{particle}), the particle and antiparticle energy, $E=- p_\mu u^\mu$, will both have the same sign at the point B  and so energy conservation will be violated around a dynamical black hole if we have pair creation around the dynamical apparent horizon. But a created particle around the point  A  can have a different sign of energy from the antiparticle (which preserves energy conservation) and so can become a long-lived actual particle. This energy calculation is a classical result.  However, we know that energy conservation will be broken in the quantum physics case on small enough scales.  From the uncertainty principle (UP)  we know that 
\be\label{eq:uncert}
\Delta x^\mu \Delta p_\nu \geq \frac{\hbar}{2} \delta^\mu_\nu\,\,,
\ee
which is  the covariant form of the UP presented in Appendix A (Eqn.(\ref{etf})).
The particle/anti-particle picture can be applied in the dynamic case using this uncertainty principle, which shows how particles can be created around a black hole by vacuum fluctuations. 
In the next section we will how this 
 can be used to characterize the 
 particle creation phenomenon around a dynamical black hole.\\

We do not claim that created particle must have a special wavelength, but rather 
give a constraint on the frequency (wavelength) range that is derived from the dynamical nature of the process. 
 We do not use any specific one of the different methods often used, and particularly the tunneling method, to get our result: we just use the quantum uncertainty principle to give a constraint on the spectrum of Hawking radiation that will hold for all methods.

\section{Particle creation rate for general black holes}\label{sec:3}

In quantum physics, quantum 
fluctuations allow the creation of particle-antiparticle pairs of virtual particles. These pairs exist for an extremely short time, and then reciprocally annihilate. In some cases, however, it is possible to boost the pair apart using external energy, therefore they avoid annihilation and become actual  (long-lived) particles. It was shown in  \cite{firouzjaee-ellis15} that to have long-lived particle creation (which respects energy conservation)  around a black hole, it is necessary that one particle falls into the black hole (that is, inside the apparent horizon) and the second remains outside the apparent horizon. In addition we know that energy conservation can be violated for short times according to the UP. Now, we want to see when this principle allows that a virtual particle pair becomes an actual pair around a dynamical black hole. \\

Consider the right hand side of  Fig.(\ref{particle}). At first  the black holes radius grows, but finally it becomes isolated \cite{isolated} at point A. Note that here the \textit{event horizon} is the causal boundary (a null surface) for the region from which the particle can escape from falling into the black hole.  The only case when particle creation can occur is when  the antiparticle falls into the apparent horizon so as to have negative energy, and the particle gets to infinity (or to a distant observer) from outside the event horizon.  According to the UP (\ref{eq:uncert}), if a created particle near the  horizon (in the simple case, a  photon which moves on a  null geodesic at the point B) is supposed to be seen at future infinity and the related antiparticle falls  inside the apparent horizon, the particle --- antiparticle distance must satisfy 
\be \label{deltatau}
\Delta x > \Delta x_0 |_{dfree}
\ee
where  $\Delta x_0$ is the distance between event and apparent horizon seen by a  Free Falling observer released on the far distance. A \textit{dfree} observer is defined as a geodesically moving observer who starts from rest at a large distance, which is outside the apparent horizon because we are considering the dynamical case with infalling positive density radiation. Here an observer is defined to be `at rest' if they are such that the area coordinate $R$ is instantaneously constant for them at the time considered: \begin{equation}\label{FFH}
	dR/d\tau = 0\,.
	\end{equation} 
	 This is possible when the surfaces $R = const$ are timelike, which will be true at the event horizon or large distance which is outside the apparent horizon in the dynamic case with infalling matter. This observer is in fact a  Kodama observer \cite{Kodama:1979vn}. It is shown in Appendix B this vector is timelike outside the apparent horizon, even though the apparent horizon itself is spacelike if matter is falling in.\\

Then $\Delta \tau= \Delta x_0 $ is the time that a created particle takes to move from the event horizon to the apparent horizon, in the \textit{dfree} frame. 
As it is outside the event horizon, 
 this particle can be seen from  outside. Here we assume that the time scale for a free falling observer to pass between the two horizons is shorter than the time scale of the apparent horizon evolution. \\

According to the UP, the allowed range of momentum for these created particles is \be
\Delta p < \Delta p_0 =\frac{\hbar}{\Delta x_0}.
\ee
In term of the energy-time UP, we get 
\be
0< \Delta E < \Delta E_0 =\frac{\hbar}{\Delta \tau}\,.
\label{energy2}
\ee
 On defining $w:= \Delta E/\hbar$, the relation
\be
0< w < \frac{1}{\Delta \tau}=:w_{up}
\label{up fre bound}
\ee  
determines the allowed energy spectrum of radiation from a dynamical black hole in terms of the massless particle frequency (the energy must be positive because of (\ref{energy})). To have creation of a real particle as seen by the \textit{dfree} observer 
the energy should be in this interval, where all quantities are measured relative to the \textit{dfree} observer.\\

There is a similar result for the Schwinger effect \cite{Aitchinson}. To have actual particles arising from a virtual pair, the probability to find one particle
at the distance $\Delta x_0$ from another one is proportional to $exp(-\frac{\Delta x_0}{\lambda})$, where $\lambda=\frac{\hbar}{mc}$ is the Compton length for a particle of mass $m$. In the case that the particles are photons, we have 
 photons created if $\lambda_\nu > \Delta x_0$ or $w<\frac{1}{\Delta x_0}$, which is similar to equation (\ref{energy2}).\\

There are two further 
constraints on the black hole radiation spectrum \cite{Visser:2014ypa}. The first one comes from the adiabatic condition
\be
w_{min}=\sqrt{|\dot{\kappa}|}\lesssim w
\ee
where $\kappa$ is the surface gravity defined by \cite{visser03} in the case of an evolving black hole, which says that the spacetime geometry must be slowly
evolving on  the time-scale set by the frequency of the Hawking photon. The second one is related to a phase space effect: 
\be \label{def1}
w \lesssim \frac{ m^2 - m_{extremal}^2}{2m}=:w_{ps}
\ee 
where $m_{extremal}$ is the extremal mass for charged or rotating black holes. This ultraviolet cutoff says that the emitted photon energy can never be greater than the available mass energy.\\


To calculate these spectral effects on the total luminosity of the black hole, we consider a   Planck-distributed flux of Hawking-like radiation which satisfies the adiabatic condition given in equation (6.26) of the first reference in \cite{visser10}: namely
\begin{equation}
\dfrac{|\dot{\kappa}(u_*)|}{\kappa^2} \ll 1,
\end{equation} 
where $\kappa=-\frac{\ddot{p}(u)}{p(u)}$ with $U=p(u)$ which $U$ and $u$ are affine parameters at future and past infinity respectively. From  \cite{Visser:2014ypa}, when we do not take the uncertainty principle into account, the total luminosity of the black hole radiation can be written in the form
\be
L=\frac{1}{2\pi} \int_{w_{min}}^{w_{ps}} dw \frac{\hbar w}{e^{\hbar w/T}-1}f(w)
\ee
with $f(w)$ coming from a grey-body factor which includes backscattering, and the temperature is \begin{equation}
T=\frac{\hbar \kappa}{2 \pi}
\end{equation} where $\kappa$ is the surface gravity . Considering the allowed range of the energy spectrum (\ref{energy2}), the radiation luminosity around the dynamical horizon becomes
\be\label{LUP}
L_{UP}=\frac{1}{2\pi} \int_{w_{min}}^{w_{max}} dw \frac{\hbar w}{e^{\hbar w/T}-1}f(w)
\ee
where 
 \begin{equation}\label{def2}
 w_{max}= Minimum \{w_{up}, w_{ps}\}.
 \end{equation} Note that equations (9)-(14) all hold in the \textit{dfree} frame. We can define another useful quantity in this frame: 
\be \label{probability}
\mathcal{P}=\frac{\int_{w_{min}}^{w_{max}} dw \frac{\hbar w}{e^{\hbar w/T}-1}f(w)}{\int_{0}^{w_{ps}} dw \frac{\hbar w}{e^{\hbar w/T}-1}f(w)}
\ee
We call $\mathcal{P}$ the \textit{probability creation rate} for a dynamical black hole. This quantity says  by what fraction the luminosity will be decreased when considering a dynamical black hole relative to a stationary black hole with the same mass. The effect of the UP (\ref{energy}) is via the quantity $w_{max}$,  which gives an ultraviolet cutoff of the radiation spectrum. It is clear from (\ref{def1}) and  (\ref{def2}) that  $0 \leq \mathcal{P} \leq 1$. If $\Delta E_0 = 0$, then $\{\mathcal{P} = 1\} \Rightarrow \{L_{UP} = L\}$. Another feature of 
 this creation rate function is that it is proportional to $\hbar$, which indicates that this is a quantum effect. In the classical limit $\hbar \rightarrow 0$ there  is no particle creation. This quantity is calculated relative to a far distant observer. For this observer, the thermal form of the luminosity is kept. Note that for calculating the quantity $w_{up}$, we need to calculate the distance between two horizons as seen by a \textit{dfree} observer on the horizon.\\

For highly dynamical black holes, the distance between the apparent and event  horizons becomes greater and  $w_{up}$ will be smaller, hence we have less radiation flux: $\Delta \tau \gg 1 \Rightarrow \mathcal{P}\ll 1$. In contrast if the radiation influx into the black hole becomes smaller, the distance between the two horizons becomes smaller and $w_{up}$ will be greater. Then $\Delta \tau \simeq 0 \Rightarrow \mathcal{P} \simeq 1$. \\

Note that, a far distant freely falling observer \textit{dfree} is defined for well-defined backgrounds such as a black hole embedded in  asymptotically Minkowski or FLRW spacetimes.
If $R$ is again the area distance coordinate 	and $\tau$ is the proper time of the \textit{dfree} observer, the reference frame \textit{dfree} is defined as that for which 
 	\begin{equation}\label{eq16}
 	\frac{dR}{d\tau}|_{r \rightarrow \infty}=0.
 	\end{equation}
 This is uniquely defined because unlike Minkowski spacetime, these spacetimes are not invariant under the Lorentz group (although they are asymptotically flat).   One could use Gullstrand-Painlev\'{e}
  coordinates to describe the free falling observers at large distance. 
For a more generic backgrounds there is no unique method for defining the reference frame in the curved spacetime. \\

 We fix the frequency relative to this  observer \textit{dfree} by   $w_{up}=w_{dfree}$ as the observed frequency. The standard method of deriving Hawking radiation uses the same reference for the thermal radiation frequency. In the Bogoliubov transformation method \cite{Haw75} we choose the observer vacuum  at a large  distance from Schwarzschild black hole and read the thermal radiation frequency from it. For reading the thermal radiation from the quantum stress tensor we use the same distant observer to read the luminosity of thermal radiation \cite{firouzjaee-ellis15}. Since this observer will not see the gravitational time dilation and velocity Doppler shift effects when the particle is released, the intrinsic frequency of the created particle is equal to the frequency measured by this observer at that time.\\
 
   Now we want to see how the particle goes to infinity. An arbitrary observer near the horizon will find that the created particle frequency near the horizon is given by \begin{equation}
w_H=(u_\mu k^\mu)_H
\end{equation}
where $k^\mu$ is the null particle four vector, $H$ means the horizon position, and $u^\mu$ is the observer four vector. This frequency will be changed by the gravitational time dilation and velocity Doppler shift relative to the far distant observer. As shown in Fig.\ref{particle}, different freely falling observers will see different frequencies for the received radiation frequency.\\
 
In the spherically symmetric case where the metric take the form
\be
ds^2=-a(t,r)^2 dt^2 + b(t,r)^2 dr^2+ R(t,r)^2 d\Omega^2,
\ee
let us transform the metric to the areal coordinate, $R(t,r)$. With this transformation, $dR=R'dr+\dot{R}dt$, so 
\be
ds^2=-(a(t,r)^2+\dfrac{b(t,r)^2 \dot{R}^2}{R'^2}) dt^2 +2 \dfrac{b(t,r)^2 \dot{R}}{R'^2} dR dt + \dfrac{b(t,r)^2}{R'^2} dR^2+ R(t,r)^2 d\Omega^2.
\ee
We know that a free-falling observer does not detect anything abnormal when falling into the black hole. We can transform the time of this metric to the proper time of a free falling observer i.e $\tau=t+C(t,R)$. Moreover, we demand that the constant time slice be flat (this means that we choose $C$ so  that the coefficient of $dR^2$ becomes 1).
As a result, we get the following form of the metric relative to this free-fall observer: 
\be \label{dfree-met}
ds^2=-p(\tau,R)^2 d\tau^2 + q(\tau,R)^2 dR d\tau+ dR^2 + R^2 d\Omega^2.
\ee

In the Schwarzschild case where $a^2=b^{-2}=(1-2m/R)$, the \textit{dfree} observer has proper time 
\be
\tau=t+C(t,R)=t-2\sqrt{2m R}-2m \ln\dfrac{\sqrt{R}-\sqrt{2m}}{\sqrt{R}+\sqrt{2m}}.
\ee
Consequently, we get the Painleve metric
\be
ds^2=-(1-2m/R)d\tau^2 + 2\sqrt{\frac{2m}{R}} dR d\tau+ dR^2 + R^2 d\Omega^2.
\ee
As is clear from this form of the metric,  the areal coordinate $R$ is the proper length relative to the proper time coordinate in the radial direction.\\

This quantity is invariant for any time and spatial (radial)
coordinate transformation, so different radial observers will see the same areal distance between the two points.
The backreacted energy due to Hawking radiation will reduce the black hole mass,    
 determined as the Misner-Sharp mass $m=\frac{1}{2}R_H$ \cite{parikh2000}, which quantifies  the black hole radius. 
For the \textit{dfree} observer, the distance between two horizons can be expressed as 
\begin{equation}
\Delta x_0 =\Delta R_H = 2 \Delta m\, .
\end{equation}
The  $d\tau=constant$ hypersurface in (21) is flat, which leads to (24). 
In the case that the black hole mass grows from $m=m_0$ to  $m_{iso}$ and becomes isolated, the free falling observer will see
 \begin{equation}
 \Delta x_0 =2 (m_0-m_{iso}).
 \label{distance-ss}
 \end{equation}

We choose the \textit{dfree} observer's frame as a standard reference frame for measuring radiation frequencies. But a different observer at different radial point will measure different frequencies. To see this change in the frequencies, we know that the gravitational time dilation and Doppler effect can be calculated by the redshift formula. As shown in the Fig.(\ref{particle}) the redshift of the frequency from a point C to point O where  the \textit{dfree} observer is located is
\be 
1+z=\dfrac{w}{w_{dfree}}=\frac{(u_\mu k^\mu)_C}{(u_{dfree}^\mu k_\mu)_O}.
\ee
Hence for an observer at point C the radiation spectrum is blueshifted with frequency $w=(1+z)w_{dfree}$. This is like the Schwarzschild case in which the standard Hawking thermal flux is measure by a distant observer at rest. Their  result will be the same as that of the \textit{dfree} obsrver, because their velocities are the same at a large distance and the redshift relation (26) depends only on the 4-velocity, not the acceleration (which will be different for them). 
Any other observer who is nearer will see a blue-shifted spectrum \cite{Chakraborty:2015nwa, observer} in the case that the observer starts off from rest.

\section{An example: the Vaidya black hole}\label{sec:vaidya}
To estimate the creation rate function for a dynamical black hole, we apply this formalism to a dynamical Vaidya metric \cite{Vaidya:1951zz} (an exact spherically symmetric solution of the Einstein equations for infalling radiation). In principal, these calculation can be extended to non-spherically symmetric cases. Note that this geometry, which exactly represents infalling radiation, will apply only until Hawking radiation starts and the outgoing flux becomes dominant.

\subsection{Ingoing Vaidya metric}
The Vaidya spacetime metric is
\ba
ds^2 &=& - \left(1- \frac{2m(v)}{r} \right) dv^2 + 2 dv dr + r^2 d \Omega^2 \label{metric}
\ea
and from the Einstein equations has stress-energy tensor 
\ba
T_{ab} &=& \frac{dm/dv}{4 \pi r^2} dv_a ~dv_b \,  \label{stressenergy}
\ea
where $m(v)$ is the mass and $d \Omega^2 = d \theta^2 + \sin^2 \theta d \phi^2$.
The outgoing, $\ell$, and ingoing null geodesic, $n$, which are cross normalized, $\ell.n=-1$, are:
\begin{align}
&{\ell} =  \frac{\partial}{\partial v}  +  \frac{1}{2}\left( 1 - \frac{2m(v)}{r} \right)   \frac{\partial}{\partial r} 
\; \;  \label{NullVectors} \\
&n = - \frac{\partial}{\partial r}   \, .  \nonumber
\end{align}
Their expansions are 
\begin{align} 
\theta_\ell =  \frac{(r - 2 m)}{r^2}  \; \; \mbox{,} \; \; \theta_n =  - \frac{2}{r} \,  \, . \label{tLtN}
\end{align}
If $v^\mu=\ell^\mu - C n^\mu$ is the tangent vector to the apparent horizon, where $C$ is the evolution parameter, we find the following equation for Vaidya black holes:

\ba
C = 2 \frac{dm}{dv} \, \, . 
\ea
This shows that the parameter $C$ is proportional to the matter flux which falls into the black hole. \\

Assume that the evolution is characterized by a time-scale $\Lambda$ so that 
\ba
m(v) = m_i M \left( V \right) \, , 
\ea
where $ V=\frac{v}{\Lambda}$,  and $m_i$ is the initial mass.
To describe the near-equilibrium limit which we expect (see \cite{Booth:2010eu}) to occur at
large $\Lambda$ (in this case ``large'' means large relative to $m_i$), we set 
\ba
\Lambda = m_i L  \, . 
\ea
For the perturbative calculations  $(1/L)$ is a useful expansion parameter.\\ 

Defining a new radial coordinate $R$ by 
\ba
r = R m_i  \, , 
\ea
the Vaidya metric becomes
\ba
ds^2 = m_i^2 \left[ - \left(1 - \frac{2 M(V)}{R} \right) L^2  dV^2  + 2 L dV dR + R^2 d \Omega^2 \right] \, . \label{scaledmetric}
\ea
\subsection{The Radiation formula: Slowly evolving case}
As shown in \cite{Booth:2010eu}, up to second order in $(1/L)$,
the distance between the event horizon and apparent horizon can be obtained from
\begin{align}
R_{EH} \approx  2M(V) \left( 1+  \frac{4 \dot{M}}{L} + \frac{32 \dot{M}^2 + 16 M \ddot{M} }{L^2}  \right)  \label{EHcandidate} \,.
\end{align}

This expansion ensures that the black hole horizon is a  slowly evolving horizon (SEH) \cite{boothslow}. Here the dot indicates a derivative with respect to $V$. The distance between the event horizon and apparent horizon  for the \textit{dfree} observer is
\begin{equation}
\Delta x= 2 m(v)\left(  4\frac{dm}{dv} + 32 (\frac{dm}{dv})^2 + 16 m \frac{d^2m}{dv^2} \right). \label{horiozon distance}
\end{equation}
From (\ref{energy2}) the allowed energy spectrum for this black hole is \begin{equation}
\Delta E_0=\frac{\hbar}{ 2 m(v)\left(  4\frac{dm}{dv} + 32 (\frac{dm}{dv})^2 + 16 m \frac{d^2m}{dv^2}  \right)}
\end{equation} and the ultraviolet frequency for $L_{UP}$ is
\be
\label{upfre}
w_{up}=\frac{1}{ 2 m(v)\left( 4\frac{dm}{dv} + 32 (\frac{dm}{dv})^2 + 16 m \frac{d^2m}{dv^2} \right)}.
\ee
Therefore, by determining this frequency we can determine the frequency spectrum range and then the particle creation rate from (\ref{LUP}). Note that the equation (\ref{horiozon distance}) (which characterizes the distance between the two horizons in the slowly evolving horizon case) does not work for dynamical black holes. It was derived on the assumption  that $\dot{m}\ll 1$.

\section{Numerical values: galactic black holes }

For a galactic black hole with mass $10^6 M_{\odot} $ with accretion rate of about $\frac{dm}{dv}=10^3 \frac{M_{\odot}}{year}$ and assuming $ m \frac{d^2m}{dv^2} \ll \frac{dm}{dv} $, 
in SI units
\be
\Delta E_0 \approx \frac{ \hbar c^6}{G^2}\frac{1}{m \frac{dm}{dt}(1+\frac{8G }{c^3}\frac{dm}{dt})} \sim 10^{-26} J,
\ee
so we have \begin{equation}\label{eq11}
w_{up}=10^8 Hz.
\end{equation} In SI units,  
\be\label{eq12}
w_{min}=\frac{c^{3/2}}{G^{1/2}}\frac{\sqrt{\dot{m}}}{m}=10^{-16} Hz
\ee
 and 
\be
 w_{ps}=\frac{m c^2}{\hbar}=10^{18} Hz
\ee
Since the $w_{ps} > w_{up}$, the ultraviolet frequency spectrum will be \begin{equation}
w_{min} \lesssim w \lesssim w_{max}=w_{up}
\end{equation} with $w_{min}$ given by (\ref{eq12}) and $w_{up}$ by (\ref{eq11}). With this frequency spectrum, the numerical value for the probability creation rate (\ref{probability}) of this black hole is
$$\mathcal{P} \sim 1.$$
This is not strange because the matter flux $\dot{m} \ll 1$. 
\\

To present the horizon phase in terms of  geometrical parameters, we need to define an invariant parameter. The $C$ parameter is not invariant under changing the null geodesic parameter,  so we have to make an invariant quantity from this function and calculate it for the black hole apparent horizon. This new quantity is
\be
\bar{C}=C \theta_n^2 R_H^2,
\ee
and we get \begin{equation}
\bar{C}=4 C= 8 \frac{dm}{dv}
\end{equation} for the Vaidya model. The numeric value of $\bar{C}$ for a galactic black hole is \begin{equation}
\bar{C}=8\frac{G}{c^3} \frac{dm}{dt} \sim 10^{-8}
\end{equation} which verifies that the black hole  horizon is a slowly evolving horizon. \\

As a result, we consider Hawking radiation for  dynamical black holes that have slowly evolving horizons. For  all values of the matter flux such that $\bar{C} \sim \dot{m} \ll 1$ and we have a slowly evolving horizon, for the probability creation rate (\ref{probability}) we get
\be
\mathcal{P}|_{SEH} \sim 1 .
\ee

Now let us see how the value $w_{up}$  compares with the two other frequency cutoffs. According to  equation (\ref{upfre}), more matter flux leads to smaller  
$w_{up}$ and less matter flux gives a greater value for $w_{up}$. Now let us see what happens if $w_{up}=w_{min}$ for a black hole with mass $10^6 M_{\odot} $. To satisfy this condition the matter flux must be 
\be
\frac{dm}{dt}=10^{11} \frac{M_{\odot}}{year}.
\ee
 Usually, it is not possible to have a black hole with such a huge flux. As a result, in practice $w_{up} > w_{min}$. \\
 
In addition, the matter flux of this black hole in the case $w_{up}=w_{sp}$
gives
\be
\frac{dm}{dt}=10^{-75} \frac{M_{\odot}}{year}.
\ee
 It is inconceivable to have cosmological black holes in such an isolated phase.
As a result, in practice for cosmological black holes we have 
\be
w_{min} < w_{up} < w_{sp}.
\ee
The final conclusion is that the usual galactic mass black holes in a cosmological context will be slowly evolving and will emit thermal radiation, but the temperature is below the CMB (Cosmic Microwave Background) thermal temperature so it will not be visible. As an aside, the black hole heat capacity is negative. This means that the greater the black hole mass, the less the Hawking temperature. Therefore we cannot apply the classical thermodynamic  scenario to them. Namely, if heat comes from the hotter CMB radiation to the black holes, the CMB flux grows the black hole mass \cite{firouzjaee-ellis14} and the black hole temperature will decrease. Hence, the black hole becomes  colder. There is not a classical thermodynamic interpretation as in classical general relativity. Black hole radiation connects  geometrically to other heat fluxes like the CMB.\\

The assumed accretion rate is much greater than that normally considered astrophysically
	reasonable, and is vastly super-Eddington. Therefore, the fact that
	the dynamic effects are small in this case is evidence that deviations
	from the stationary black hole Hawking radiation are negligible for
	any astrophysical black hole in the present universe. We show here that the Hawking radiation luminosity for astrophysical black hole is very close to the stationary  black hole luminosity.

\section{Dynamic black holes: Primordial black holes}

To have a good intuition for the spherical dynamical black hole particle creation rate in the case of dynamic black holes, consider a dynamical black hole such that its  initial apparent horizon radius is $r_i = 2 m_i$ where $m_i$ is the initial Misner-Sharp mass \cite{firouzjaee-penn} and it's radius grows (due to the infalling matter flux) till it becomes isolated  and its  final radius becomes 
 \begin{equation}
r_f = 2 (m_i + \delta m)=r_i + 2\delta m.
\end{equation} 
As shown in equation (\ref{distance-ss}) the distance between the two horizons relative to the \textit{dfree} observer is $\Delta x_0 =2\delta m $. Hence, the frequency $w_{up}$ can be obtained from equation (\ref{up fre bound}). \\

We define the new parameter $n$ as \begin{equation}
n= \frac{\Delta x_0}{r_i}= \frac{r_f-r_i}{r_i}
\end{equation} which shows how much the black hole horizon grows relative to its initial radius. To get the order of magnitude for the probability creation rate (\ref{probability}) we approximate the spectrum as a Planckian spectrum, as is also proposed in the tunneling method \cite{Vanzo:2011wq} in calculating the luminosity order of magnitude. Neglecting  $w_{min}$, we find  the probability creation rate (\ref{probability}) for different numbers $n$  as shown  in Table (\ref{prob table}). \\
\begin{table}[h]
	\begin{tabular}{ |p{2cm}||p{1cm}|p{1cm}|p{1cm}|p{1cm}|p{1cm}|  } 
		\hline
		\multicolumn{6}{|c|}{Creation rate} \\
		\hline
		$n=\frac{r_f - r_i}{r_i}$ & $5$ & $10$ & $15$ & $20$ & $25$  \\
		\hline
		$\mathcal{P}$ & $0.99$ & $0.75$ & $0.46$ & $0.28$ & $0.18$  \\ 
		\hline
	\end{tabular}\\
	\caption{$n$-Creation rate.}
	\label{prob table}
\end{table}

This table shows an interesting result: that for a dynamical black hole such that its final radius becomes $25$ times greater than its initial radius, 
the luminosity of the particle creation rate dies off by a factor $\mathcal{P} = 0.18$. This result is compatible with the intuition from Fig.(\ref{particle}) that bigger $n$ means bigger growth of the horizon radius, and less created particles around $r_i$ at point B can escape to outside of $r_f$ at point A. \\

This result changes the expected primordial black hole radiation when they are located in the hot soup of the very early radiation dominated era. It is probable that primordial black holes that were created in the early universe eat this much matter flux, and their initial radius grows by this amount.
The UP frequency visibility, $w_{up}$,   is another observational effect in our analysis. For a Primordial black hole with mass $m=10^{15} gr$ if $n=0.1$ or $\Delta x_0 =0.1 r_i$ the upper bound frequency becomes $w_{up}= 10^{11} Hz$ which can be observable. Therefore, observing the radiating frequency cutoff can verify that this black hole was in a dynamical phase at the emission moment.

\section{Conclusion}

As this is a quantum effect, to understand the nature of particle creation near a general black hole we need to adopt a radiation scenario that takes into account 
quantum fluctuations. The essence of quantum fluctuations is based in Heisenberg's uncertainty principle (UP) at the quantum level. We have used this principle to study the nature of Hawking radiation near a dynamical black hole. Applying the UP allows one to set limits on the energy spectrum of the radiation. The radiation  spectrum has a dependency on the distance  $\Delta x_0$ between the apparent and event horizons which is measured by the \textit{dfree} observer. The new proposed spectrum (\ref{LUP}) will reduce to the standard spectrum when the geometry becomes stationary. We also defined the \textit{probability creation rate} (\ref{probability}) to show by  how much black hole radiation will be decreased  due to the dynamical nature of a black hole. Note that we did not claim that particle creation takes place just on the horizon. We determined the minimum distance between the horizons allowing pairs to be real in the dynamical case. From that we derived an upper bound for the radiation frequency cut off. \\

To apply this to  known black hole models, first we took the Vaidya model where we can explicitly calculate the distance between the two horizons. But it is in an adiabatic phase, and has a slowly evolving horizon. We have shown that the probability creation rate for such a Vaidya black hole which has a slowly evolving horizon is $\mathcal{P} \sim 1$. This verifies our last work \cite{firouzjaee-ellis15}, where we inferred that black holes with slowly evolving horizons emit Hawking radiation. This result also is compatible with the 
thermodynamics principle that says temperature is meaningful only in the case of equilibrium or near equilibrium systems.\\

 To study dynamical models, we need to make numerical simulations.  
It is suggested by this investigation that a black hole  starts emitting radiation when it becomes slowly evolving. But our spectrum discussion concerns general black holes, and will apply in that case as well as the dynamic case. In addition, to get some intuition for the probability creation rate of a dynamical black hole, we calculated this quantity for a spherical black hole such  that its apparent horizon grew several times due to accretion. As shown in  the Table (\ref{prob table}), the probability creation rate will not be negligible in the case that its horizon grows $10$ times greater. This seems to be an important result for  primordial black hole radiation. We have inferred that this frequency cut off, $w_{up}$, can be observable for dynamical primordial black holes, and we can consider it to be a parameter which shows the primordial black hole growth at the emission moment. Since these black holes were created in the radiation dominated era when they were surrounded by a hot soup, these black holes can grow  substantially. This paper confirms our last paper's results \cite{firouzjaee-ellis14} that we have to revise the primordial black hole radiation formula to take this into account.\\

The case of a primordial black hole considered here does exhibit a breakdown of the quasi-stationary approximation,
 but one might assume that Hawking radiation is totally negligible compared with the thermal
 radiation of the early universe and classical gravitational radiation
 associated with the black hole formation process. There may be no observational consequences of the result. 
  However our result challenges the dynamical primordial black hole radiation result which is mentioned in the literature \cite{PagHaw76}.  To check this result one needs to quantify the Hawking radiation flux and compare it with the incoming flux of thermal radiation of the early universe. The scope of this paper is not to consider in detail primordial or astrophysical BHs, it is to consider general principles at work. Table I is just an example, but is not applied to any specific simulation.

\appendix
\section{Covariant Uncertainty Principle}
Although there is not a full relativistic version of quantum mechanics, we can generalize some quantum mechanics aspects to the curved spacetime case. Consider a test particle (it does not affect the spacetime curvature) which has passive mass  $m$. The four-momentum of this particle is $P^\mu = m v^\mu$ where $v^\mu$ is the particle four-velocity. 
An energy-momentum tensor follows from a Lagrangian for this particle, and 
 this energy-momentum tensor gives the four-momentum vector measured by an  observer with four-velocity $u^\mu$ as $P^\mu =u_{\nu} T^{\mu \nu}$. We  define the Hermitian operator $\hat{P}^\mu $ which corresponds to the energy-momentum vector which generates the infinitesimal change, $dx^\mu$, in the curved spacetime as
\be
|\psi(x_0+dx^\mu)\rangle = e^{-i dx_\mu \hat{P}^\mu/\hbar} |\psi(x_0)\rangle 
\ee
A path in the Hilbert space 
is generated by the operator  $\hat{P}^\mu = -i \hbar \nabla^\mu$. The proper time (invariant distance) for this infinitesimal transformation becomes 
\begin{equation}
	ds^2=g_{\mu \nu} dx^\mu dx^\nu
\end{equation}
with the unit orientation $n^\mu= \frac{dx^\mu}{ds}$.
Using the metric signature $(-,+,+,+)$ we define the covariant canonical commutation relations
$$[X^\mu, \hat{P}^\nu]= i \hbar g^{\mu \nu},
$$
where $X^\mu$ measures the distance of the particle from the origin. One can define the projection matrix $h^{\mu \nu}=g^{\mu \nu} +u^\mu u^\nu$ which projects tensor quantities to the hypersurface which is orthogonal to $u^\mu$.
Any measurement of space and momentum has an uncertainty relation  
\be
\langle h^\mu_\rho dx^\rho \rangle  \langle h_{\nu \alpha} \hat{P}^\alpha \rangle \geq \frac{\hbar}{2} h^\mu_\nu
\ee
relative to the observer with 4-velocity $u^\mu$. Similarly, one can infer that the time-energy measurement has the uncertainty relation:
\be
\label{et}
\langle u_\rho dx^\rho \rangle  \langle u_\alpha \hat{P}^\alpha \rangle \geq \frac{\hbar}{2}.
\ee
Using the equivalence principle the free-falling observer see these uncertainty relations as a simple form
\be \label{etf}
\Delta x^\mu \Delta p_\nu \geq \frac{\hbar}{2} \delta^\mu_\nu.
\ee
where $\Delta x^\mu=(\Delta t,\Delta x^i)$ and $\Delta p^\mu=(\Delta E,\Delta p^i)$ with $i=1,2,3$.
\section{Kodama vector in the spherically symmetric case}
Take
a collapsing ideal fluid within a compact spherically symmetric
spacetime region described by the following metric in comoving
coordinates $(t,r,\theta,\varphi)$:
\begin{equation}
ds^{2}=-e^{2\nu(t,r)}dt^{2}+e^{2\psi(t,r)}dr^{2}+R(t,r)^{2}d\Omega^{2}.
\end{equation}
Assuming the energy momentum tensor for the perfect fluid in the
form
\begin{eqnarray}
T^{t}_{t}=-\rho(t,r),~~T^{r}_{r}=p_{r}(t,r),\nonumber\\~~T^{\theta}_{\theta}=
T^{\varphi}_{\varphi}=p_{\theta}(t,r)=w \rho(t,r),
\end{eqnarray}
with the weak energy condition
\begin{equation}
\rho\geq0~~\rho+p_{r}\geq0~~\rho+p_{\theta}\geq0,
\end{equation}
where $w$ is constant. The Einstein equations give,
\ba \label{gltbe2}
\rho=\frac{2M'}{R^{2}R'}~,~~p_{r}=-\frac{2\dot{M}}{R^{2}\dot{R}},
\ea
\begin{equation}
\nu'=\frac{2(p_{\theta}-p_{r})}{\rho+p_{r}}\frac{R'}{R}-\frac{p'_{r}}{\rho+p_{r}},
\end{equation}
\begin{equation}
-2\dot{R}'+R'\frac{\dot{G}}{G}+\dot{R}\frac{H'}{H}=0,
\end{equation}
where
\begin{equation}\label{ms1}
G=e^{-2\psi}(R')^{2}~~,~~H=e^{-2\nu}(\dot{R})^{2},
\end{equation}
and $M$, the Misner-Sharp mass,  is defined by
\ba 
\label{gltbe3} G-H=1-\frac{2M}{R}. 
\ea
It can be shown  \cite{firouzjaee-penn} that if a perfect fluid collapses the apparent horizon will form at $R=2M$. Consider a two-sphere with $R=constant$. We can define 
 timelike and spacelike normal vectors to this compact surface. Consider the radial  vector \begin{equation}
K^\mu=e^{-\nu-\psi}(R',-\dot{R},0,0)
\end{equation} which is called the Kodama vector. The amplitude of this vector is 
\be
|K|^2=-K_\mu K^\mu=e^{-2\nu-2\psi}(e^{2\nu} R'^2-e^{2\psi} \dot{R}^2) \, .
\ee
Using equation (\ref{ms1}) and equation  (\ref{gltbe3})
this equation gives
\be
-K_\mu K^\mu=1-\frac{2M}{R}. 
\ee
It can thus be easily seen that the Kodama vector outside the apparent horizon ($R>2M$) is timelike; inside the apparent horizon ($R<2M$) is spacelike; and on the apparent horizon ($R=2M$) is null. On the other hand, we know that the event horizon is located outside the apparent horizon. As a result, the Kodama vector is timelike on the event horizon.\\

{\bf Acknowledgments:} We thank the Oxford University Physics Department, and particularly John Miller and Pedro Ferreira, for hospitality. We thank John Miller, Ilia Musco, Abasalt Rostami and Valerio Faraoni for useful discussions.\\

GE thanks the National Research Foundation (South Africa) and the University of Cape Town Research Committee for financial support.\\

\end{document}